\documentclass[twocolumn,showpacs,amssymb,aps,prl]{revtex4}

\usepackage{epsfig}

\newcommand{\bra}{\langle}
\newcommand{\ket}{\rangle}

\newcommand{\Tr}{\hbox{Tr}}

\newcommand{\be}{\begin{equation}}
\newcommand{\ee}{\end{equation}}
\newcommand{\bea}{\begin{eqnarray}}
\newcommand{\eea}{\end{eqnarray}}
\newcommand{\bean}{\begin{eqnarray*}}
\newcommand{\eean}{\end{eqnarray*}}

\newcommand{\txt}{\textstyle}
\newcommand{\C}{{\cal C}}

\newcommand{\nn}{\nonumber}

\newcommand{\half} {{\txt {1\over 2}}}

\begin{document}                                                

\title{Transport coefficients from the 2PI effective action}

\author{Gert Aarts}
\author{Jose M.\ Mart{\'\i}nez Resco}
\affiliation{Department of Physics, The Ohio State University, 
Columbus, OH 43210, USA}
\date{March 25, 2003}

\begin{abstract} 

We show that the lowest nontrivial truncation of the two-particle
irreducible (2PI) effective action correctly determines transport
coefficients in a weak coupling or $1/N$ expansion at leading
(logarithmic) order in several relativistic field theories. In particular,
we consider a single real scalar field with cubic and quartic interactions
in the loop expansion, the $O(N)$ model in the 2PI-$1/N$ expansion, and
QED with a single and many fermion fields. Therefore, these truncations
will provide a correct description, to leading (logarithmic) order, of the
long time behavior of these systems, i.e.\ the approach to equilibrium.
This supports the promising results obtained for the dynamics of quantum
fields out of equilibrium using 2PI effective action techniques.

\end{abstract}


\maketitle


{\em Introduction.}
 Recent developments in heavy-ion collisions and cosmology have spurred
the theoretical understanding of the dynamics of quantum fields out of
equilibrium. In particular, the thermalization of quantum fields is a
subject of both fundamental and practical relevance. For quantum fields
far from equilibrium, promising results have been obtained from a
systematic use of the 2PI effective action \cite{Cornwall:1974vz},
formulated along the Schwinger-Keldysh contour. While the basic
formulation of this approach is well-known \cite{Calzetta:1986cq} (see
Refs.\ \cite{Blaizot:1999ip} for recent applications in equilibrium), the
recent progress has been the numerical solution of the resulting evolution
equations for the (one- and) two-point functions without any further
approximation. This allows one to go far from equilibrium and describe,
e.g., the emergence of quasi-particles in a completely self-consistent
way. This program has been carried out for a single scalar field with
quartic self-interactions using a three-loop expansion in $1+1$ dimensions
\cite{Berges:2000ur}, for the $O(N)$ model using the 2PI-$1/N$ expansion
\cite{Aarts:2002dj} in $1+1$ \cite{Berges:2001fi,Mihaila:2000sr} and $3+1$
dimensions \cite{Berges:2002cz}, and recently also for a chirally
invariant Yukawa model in $3+1$ dimensions \cite{Berges:2002wr}. The
(mostly numerical) results obtained so far suggest that the lowest nontrivial
truncation beyond the mean-field approximation is necessary and sufficient
to describe in one formalism both the dynamics far from equilibrium as
well as the subsequent equilibration.

A necessary requirement for any method to successfully describe
nonequilibrium field dynamics is that it encompasses the correct long-time
behavior. Close to equilibrium, the evolution of the system on long time-
and length scales is characterized by transport coefficients, which have
been computed at leading order in a weak coupling or $1/N$ expansion
\cite{Jeon:if,Arnold:2000dr,Moore:2001fg}. In order to assess the validity
of truncations of the 2PI effective action, it is therefore crucial that
transport coefficients obtained within the 2PI formalism agree, in the
weak coupling limit, with those results.

In this paper, we show how the calculation of transport coefficients is
organized in the framework of the 2PI effective action. We then consider a
variety of theories and show, by comparing with results obtained
previously, that truncations currently used in far-from-equilibrium
applications include in the weak coupling or large $N$ limit the
appropriate diagrams to yield the correct result for transport
coefficients to leading (logarithmic) order. This result provides strong
support for the applicability of truncations of the 2PI effective action
to describe the equilibration of quantum fields.


{\em 2PI effective action.}
 The Kubo formula relates transport coefficients to the expectation value
of appropriate composite operators in thermal equilibrium (to be precise, 
to the imaginary part
of retarded correlators at zero momentum and vanishing frequency). The
correlator we will be interested in here is therefore of the form
\be
\label{eqO}
\bra O(x)O(y)\ket - \bra O(x)\ket\bra O(y)\ket,
\ee
with $O$ an operator bilinear in the fundamental fields. For example, for
the shear viscosity $O(x) = \pi_{ij}(x)$, where for a real scalar
field $\pi_{ij} = \partial_i\phi\partial_j\phi - \frac{1}{3}
\delta_{ij}\partial_k\phi\partial_k\phi$, and for the electrical 
conductivity $O(x) = j^i(x)$, with $j^i = \bar\psi \gamma^i\psi$. 

We first demonstrate that the 2PI effective action generates 
precisely correlators of the form (\ref{eqO}). 
 We consider the case of a real field $\phi$, coupled to a bilinear 
source $K$ (the extension to fermions is straightforward). The path 
integral is
\be
Z[K] = e^{iW[K]} = \int {\cal D}\phi\, e^{i\left(S[\phi]
+\frac{1}{2}\phi^i
K_{ij}\phi^j\right)},
\ee
with $S$ the classical action. (We use a condensed notation where latin
indices summarize space-time as well as internal indices, and integration
and summation over repeated indices is understood.) For the application we
discuss here it is not necessary to couple a source to $\phi$ itself and
we assume throughout that $\bra\phi^i\ket$ vanishes. The 2PI effective
action is defined as the Legendre transform of $W$, $\Gamma[G] = W[K] -
\frac{1}{2}G^{ij}K_{ij}$, with $\delta W[K]/\delta K_{ij} = \half G^{ij}$,
and $\delta \Gamma[G]/\delta G^{ij} = -\half K_{ij}$. The two-point
function $G^{ij} = \bra T_{\cal C}\phi^i\phi^j\ket$ is the (connected)
time-ordered two-point function along the contour ${\cal C}$ in the
complex-time plane. Our discussion will be general and we do not need to
specify the contour here; at any moment one may specialize to the
Matsubara contour or the Schwinger-Keldysh contour. 

\begin{figure}[t]
 \begin{center}
 \epsfig{figure=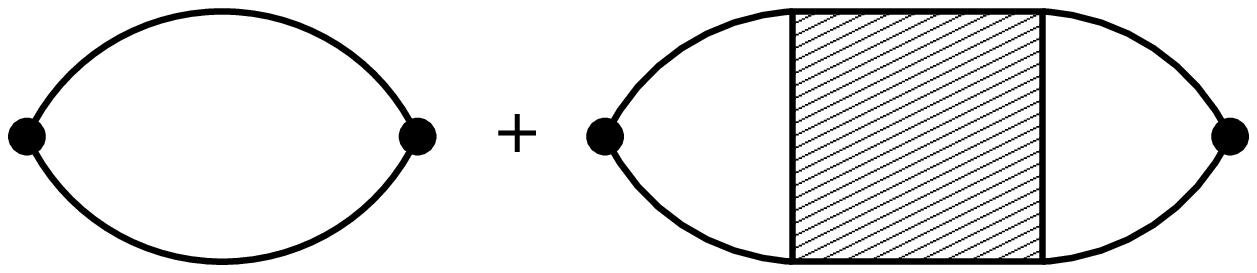,height=1.0cm}
 \end{center}
 \vspace{-0.5cm}
 \caption{Expectation value of bilinear operators (black dots) 
 from the 2PI effective action. The shaded square denotes the
 connected 4-point function (with external legs amputated).}
 \label{figexpecvalue}
\begin{center}
 \epsfig{file=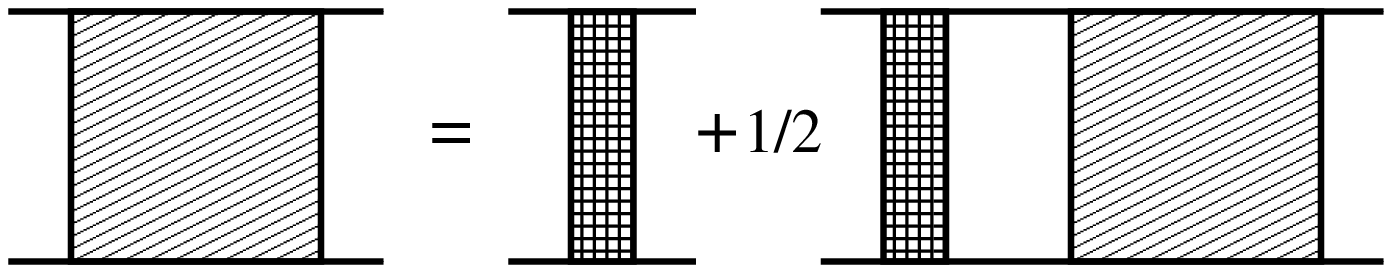,height=1.0cm}
 \end{center}
 \vspace{-0.5cm}
 \caption{Integral equation for the 4-point function, derived from
 the 2PI effective action.}
 \label{figintegral}
\end{figure}

In order to obtain a correlator of the form (\ref{eqO}), 
we differentiate $W$ twice with respect to $K$:
\bea
 \frac{\delta^2 W[K]}{\delta K_{ij}\delta K_{kl}} 
 &=& \frac{i}{4}\left[ \bra T_\C \phi^i\phi^j\phi^k\phi^l\ket
 - G^{ij}G^{kl} \right] \nn \\
 &=& \frac{i}{4}\left[ G_c^{ij;kl} + G^{ik}G^{jl} +
 G^{il}G^{jk} \right],
\eea
where $G_c^{ij;kl}$ is the usual connected 4-point function (semicolons 
separate indices with a different origin). We note 
that $W$ does not generate connected Green functions, instead it 
generates a 4-point function which, after identifying $i, j$ with $x$ 
and $k, l$ with $y$, is precisely of the form (\ref{eqO}) (see Fig.\ 
\ref{figexpecvalue}). To proceed further we remove the external legs,
$G_c^{ij;kl} = G^{ii'}G^{jj'}G^{kk'}G^{ll'}\Gamma^{(4)}_{i'j';k'l'}$, and 
concentrate on the 4-point vertex function $\Gamma^{(4)}$. Note that in a 
theory with cubic interactions $\Gamma^{(4)}$ is not 1-particle 
irreducible. 
The vertex function obeys an integral equation that can be obtained 
using standard functional relations \cite{Cornwall:1974vz}. It reads (see 
Fig.\ \ref{figintegral})
\be
\label{eqintegral}
\Gamma^{(4)}_{ij;kl} = \Lambda_{ij;kl}
+ \frac{1}{2} \Lambda_{ij;mn} G^{mm'}G^{nn'} \Gamma^{(4)}_{m'n';kl},
\ee
where the 4-point kernel follows from
\be
\label{eqlambda}
\Lambda_{ij;kl} = 2 \frac{\delta \Sigma_{ij}[G]}{\delta G^{kl}},
\;\;\;\;\;\;
\Sigma_{ij} = 2i \frac{\delta \Gamma_2[G]}{\delta G^{ij}}, 
\ee
when the effective action is written as \cite{Cornwall:1974vz}
\be
\Gamma[G] = \frac{i}{2}\Tr \ln G^{-1} + \frac{i}{2}\Tr\, G_0^{-1}(G-G_0)
+\Gamma_2[G].
\ee
Here, $iG_{0}^{-1}$ is the free inverse propagator and $\Gamma_2[G]$ is
the sum of all 2-particle irreducible (2PI) diagrams with no external legs
and exact propagators on the internal lines.  We emphasize that Eq.\
(\ref{eqintegral}) is exact and applies out of equilibrium.  The
first derivative of $\Gamma[G]$ determines the gap equation $G^{-1} =
G_{0}^{-1} - \Sigma$ in the absence of sources.

To apply the Kubo formula, we specialize to a system in thermal 
equilibrium that is invariant under spacetime translations.
In momentum space, Eq.\ (\ref{eqintegral}) then reads
\be
\label{eqintmom}
\Gamma^{(4)}(p,k) = \Lambda(p,k)
+ \frac{1}{2} \int_q \Lambda(p,q) G^2(q)\Gamma^{(4)}(q,k).
\ee
The importance of using Eq.\ (\ref{eqintmom}) in the renormalization of the 
gap equation has been emphasized recently 
\cite{vanHees:2001ik,Blaizot:2003br}. 
 Eq.\ (\ref{eqintmom}) is valid both in the imaginary-time as well as in 
the real-time formalism, where the 4-point function and kernel have a more 
complicated tensor structure \cite{Wang:2002nb}.

We note that the 4-point function appears in the integral equation with a
particular momentum configuration:  the momentum $p$ ($k$) enters and
leaves on the left (right) and the two intermediate propagators carry the
same momentum $q$. This configuration suffers therefore from pinching
poles:  when the loop momentum $q$ is nearly on-shell, the product of
propagators is potentially very large and all terms in the ladder series
may be equally important. This situation is precisely the one
that appears in the diagrammatic evaluation of transport coefficients
\cite{Jeon:if}: computing transport coefficients diagrammatically amounts
to studying an integral equation of the type (\ref{eqintmom}), specialized
to the case that the external momenta $p$ and $k$ as well as the internal
momentum $q$ are on-shell \cite{Jeon:if,Wang:2002nb}.

\begin{figure}[b]
\begin{center}
\epsfig{figure=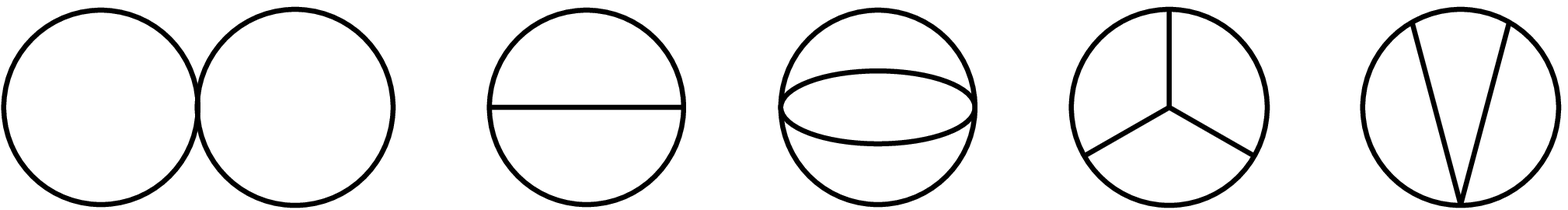,height=1.0cm}
\end{center}
 \vspace{-0.5cm}
 \caption{Contributions to the 2PI effective action in the loop expansion
up to three loops in a theory with cubic and quartic interactions.}
 \label{figgamma}
\begin{center}
\epsfig{figure=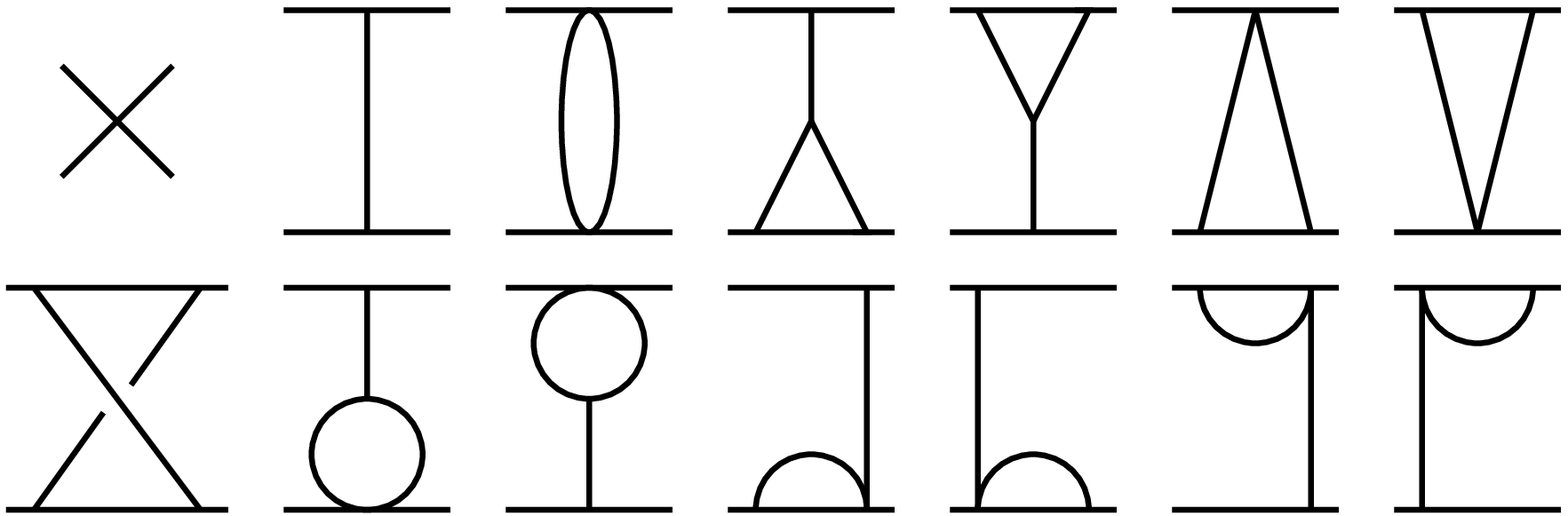,height=2.2cm}
\end{center}
 \vspace{-0.5cm}
 \caption{Zero- and one-loop skeleton kernel from the contributions to the 
 2PI  effective action shown in Fig.~\ref{figgamma}.}
 \label{figkernel}
\end{figure}

{\em Loop expansion.}
 As a first example, we consider a real scalar field with cubic and quartic 
interactions, 
\be
S = \int_x \left[\frac{1}{2}(\partial_{\mu}\phi)^2 - \frac{1}{2}m^2\phi^2 -
\frac{g}{3!}\phi^3 - \frac{\lambda}{4!}\phi^4\right].
\ee
In the loop expansion for $\Gamma_2[G]$, we include terms up to three loops 
(see Fig.~\ref{figgamma}). The self energy $\Sigma$ is obtained by cutting 
the diagrams once. Cutting the result once more yields the kernel 
$\Lambda$ in the integral equation for the 4-point function (see 
Fig.~\ref{figkernel}). 

\begin{figure}[t]
 \begin{center}
 \epsfig{figure=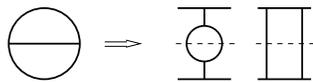,height=1.0cm}
 \end{center}
 \vspace{-0.5cm}
 \caption{One-loop perturbative kernel from the 2PI effective action at 
 two-loop order in a theory with cubic interactions. The dashed 
 line indicates how to cut the diagrams to make the connection with the 
 scattering kernel in kinetic theory.
}
 \label{figkernelpert}
\end{figure}

We now show that this truncation includes the
physics relevant for the leading-order result of the shear viscosity in
the weak-coupling limit, $\lambda \sim (g/m)^2 \ll 1$. In order to do
this, it is sufficient to show that Eq.~(\ref{eqintmom}) includes all 
diagrams in the appropriate kinematic configuration which are
known to contribute at leading order~\cite{Jeon:if}.
 We therefore specialize to on-shell momenta $p$, $k$, and $q$, where the
leading pinching-pole limit arises.
 One must then proceed
to a careful analysis of all the possible perturbative contributions to
the integral equation from the kernel $\Lambda(p,k)$. For example, the
single rung with a bare propagator does not contribute straightforwardly 
due to kinematics, but it does when either it is iterated in
the integral equation to get a one-loop kernel or a one-loop self-energy
correction is included, as depicted in Fig.~\ref{figkernelpert} (for
detailed power-counting arguments, see Ref.\ \cite{Jeon:if}). In general, 
Eq.\ (\ref{eqintmom}) will contain contributions that are
subleading in the weak-coupling limit. For instance, the four final rungs 
in Fig.\ \ref{figkernel} can be seen as vertex corrections to the single 
rung and contribute at subleading order only.
When the result of the power-counting analysis of Ref.\ 
\cite{Jeon:if} is carried over to this case, one finds that the 
leading-order contribution to the kernel can be written as an integral 
over an effective scattering kernel
$\left| \lambda + g^2\left[G_R(s) + G_R(t) + G_R(u)\right]\right|^2,$
where we write $s,t,u$ to indicate the contributions from the three
scattering channels and $G_R$ denotes the retarded Green's function. This
kernel is the square of the sum of all 2-to-2 processes in a theory with
cubic and quartic interactions, which establishes the connection with
kinetic theory \cite{Jeon:if}. It is instructive to see how this result 
is put together. Consider for a moment only the two-loop diagram with 
cubic interactions (see Fig.\ \ref{figkernelpert}). The scattering kernel 
that arises from this diagram reads explicitly (see e.g.\ 
\cite{Wang:2002nb}):
$g^4 [ \left| G_R(s)\right|^2 + \left| G_R(t)\right|^2 + \left|
G_R(u)\right|^2 ]$,
i.e., the sum of the squares of the matrix elements, however, without
interference terms. Indeed, it is easy to see that the interference terms
originate from the 3-loop diagrams. 

We conclude therefore that the 3-loop approximation of the 2PI effective
action is necessary and sufficient to yield the leading-order result for
the shear viscosity in a weakly-coupled scalar theory
\cite{Calzetta:1999ps}.

{\em $O(N)$ model.}
 We now consider a real scalar $N$-component quantum field $\phi_a$
($a=1,\ldots, N$) with a classical $O(N)$-invariant action and an 
interaction term $(\lambda/4!N)(\phi_a\phi_a)^2$. Instead of the
loop expansion, we will use the 2PI--$1/N$ expansion to next-to-leading
order (NLO), which is discussed in detail in Refs.\
\cite{Aarts:2002dj,Berges:2001fi}. 
\begin{figure}[t]
 \begin{center}
 \epsfig{figure=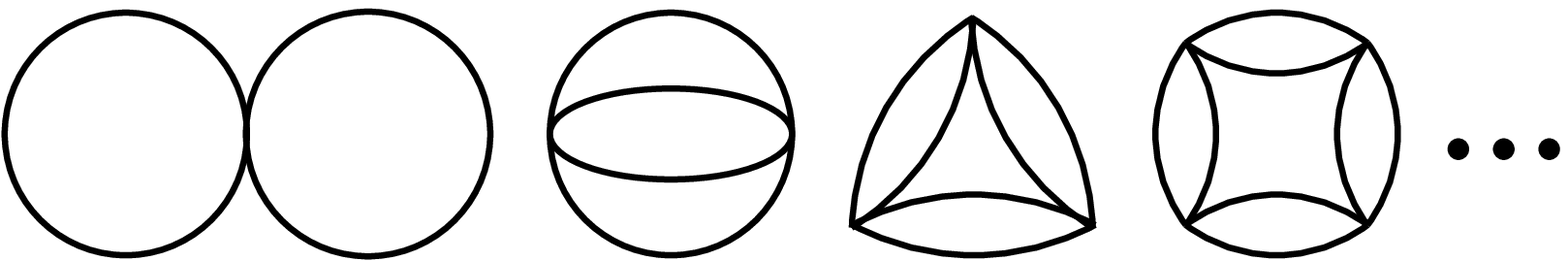,height=1.0cm}
 \end{center}
 \vspace{-0.5cm}
 \caption{Contributions to the 2PI effective action in the $O(N)$ model in 
 the 2PI--$1/N$ expansion at LO and NLO. Only the first few diagrams at NLO 
 are shown.}
 \label{figN}
 \begin{center}
 \epsfig{figure=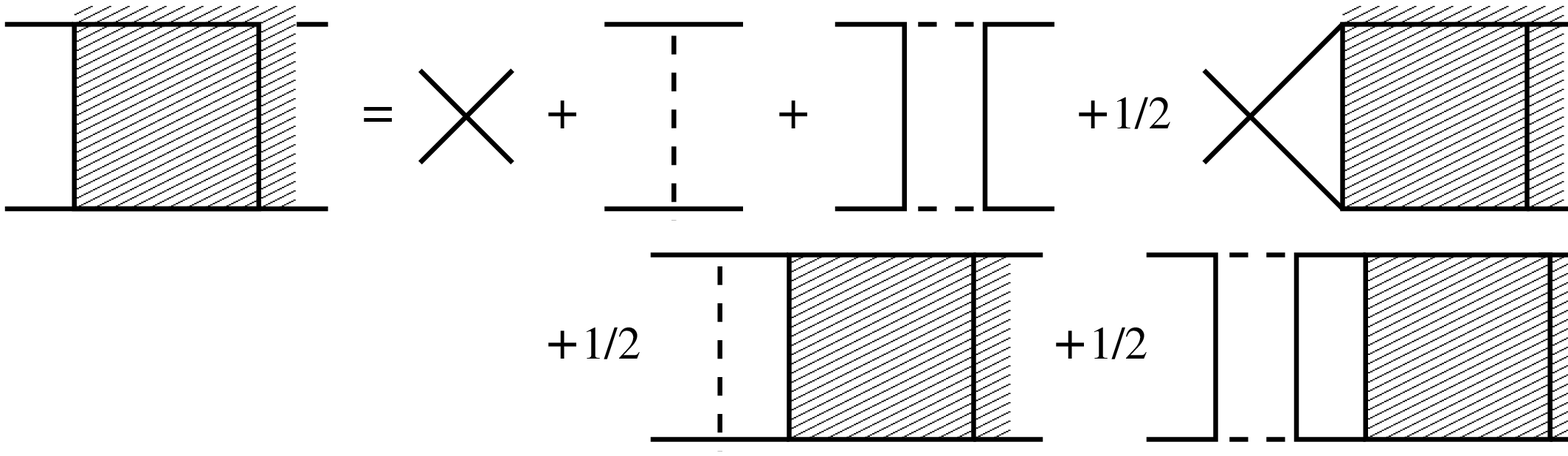,height=2.1cm}
 \end{center}
 \vspace{-0.5cm}
 \caption{Integral equation for the 4-point function in
 the 2PI--$1/N$ expansion of the $O(N)$ model at NLO.}
 \label{figintegralN}
 \begin{center}
 \epsfig{figure=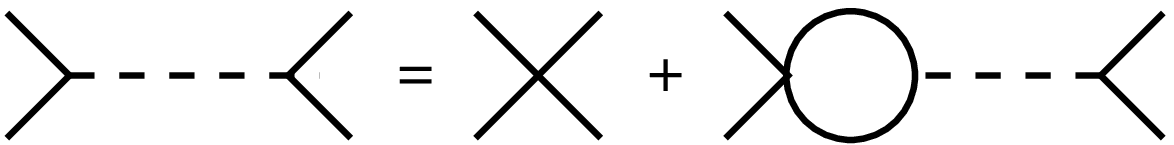,height=0.6cm}
 \end{center}
 \vspace{-0.5cm}
 \caption{Integral equation for the auxiliary correlator $D$.}
 \label{figintegralD}
\end{figure}
The 2PI part of the effective action
can be written as $\Gamma_2[G] = \Gamma_2^{\rm LO}[G] + \Gamma_2^{\rm
NLO}[G] + \ldots$, with (see Fig.\ \ref{figN})
\bea
\Gamma_2^{\rm LO}[G] &=& -\frac{\lambda}{4!N}\int_x G_{aa}(x,x)
G_{bb}(x,x),\\
\Gamma_2^{\rm NLO}[G] &=& \frac{i}{2}\Tr \ln {\rm\mathbf{B}}(G),
\eea
where ${\rm\mathbf{B}}(x,y;G) = \delta_\C(x-y) + (i\lambda/6N)
G_{ab}(x,y)G_{ab}(x,y)$
sums bubbles (which can be seen by re-expanding the logarithm).
In this case the kernel reads (see Fig.~\ref{figintegralN})
\bea
\!\!\!\!\Lambda^{\rm LO}_{ab;cd}(p,k) &=& -\frac{i\lambda}{3N}
\delta_{ab}\delta_{cd},
\\
\!\!\!\!\nn\Lambda^{\rm NLO}_{ab;cd}(p,k) &=& 
-\left[\delta_{ac}\delta_{bd} + \delta_{ad}\delta_{bc}\right]D(p-k)
\\
\!\!\!\!&& + 2 \int_q G_{ab}(p-q) D^2(q) G_{cd}(k-q), 
\eea
where we used the auxiliary correlator $D = (i\lambda/3N) 
{\rm\mathbf{B}}^{-1}$ to sum the chain of bubbles (see Fig.\ 
\ref{figintegralD}):
$D(p) = (i\lambda/3N)\left[ 1+\Pi(p)D(p)\right]$, with $\Pi(p) = 
-\frac{1}{2}\int_q G_{ab}(q)G_{ab}(p+q)$ \cite{Aarts:2002dj}.

The advantage of the large $N$ expansion employed here is that it allows
for a computation of the shear viscosity to first nontrivial order in the
$1/N$ expansion {\em without} a restriction to small $\lambda$. This is
especially pressing for applications of the $O(4)$ model to QCD
phenomenology in which the coupling constant $\lambda$ has to be taken
large. As far as we know, such a calculation of the shear viscosity has
not yet been performed. The integral equation derived here provides a
convenient starting point for that and work in this direction is currently
underway \cite{Aarts:2003}.


{\em QED.}
 As a last example we consider QED. Although for gauge theories there are
a number of issues related to the gauge symmetry (dependence on the
gauge-fixing parameter \cite{Arrizabalaga:2002hn}, Ward identities
\cite{Mottola:2003vx}) which have not been resolved completely, we think 
it is nevertheless worth analyzing to what accuracy transport coefficients
can be expected to be computed within a specific truncation of the 2PI
effective action.

\begin{figure}[t]
\begin{center}
\epsfig{figure=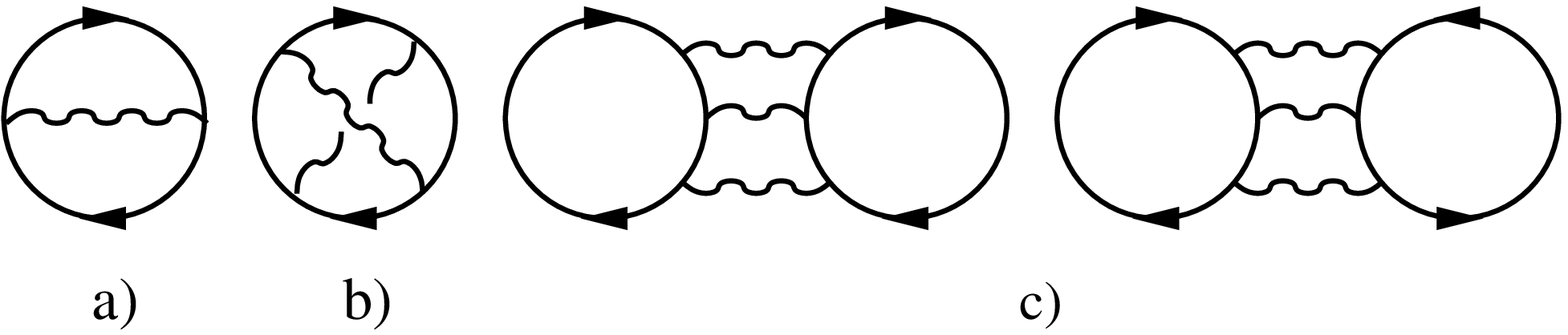,height=1.6cm}
\end{center}
 \vspace{-0.5cm}
 \caption{Contributions to the 2PI effective action in QED in the loop 
expansion with (a) 2 and (b) 3 loops or in the 2PI-$1/N$ expansion at 
(a) NLO and (b,c) NNLO. 
}
 \label{figgammaQED}
\begin{center}
\epsfig{figure=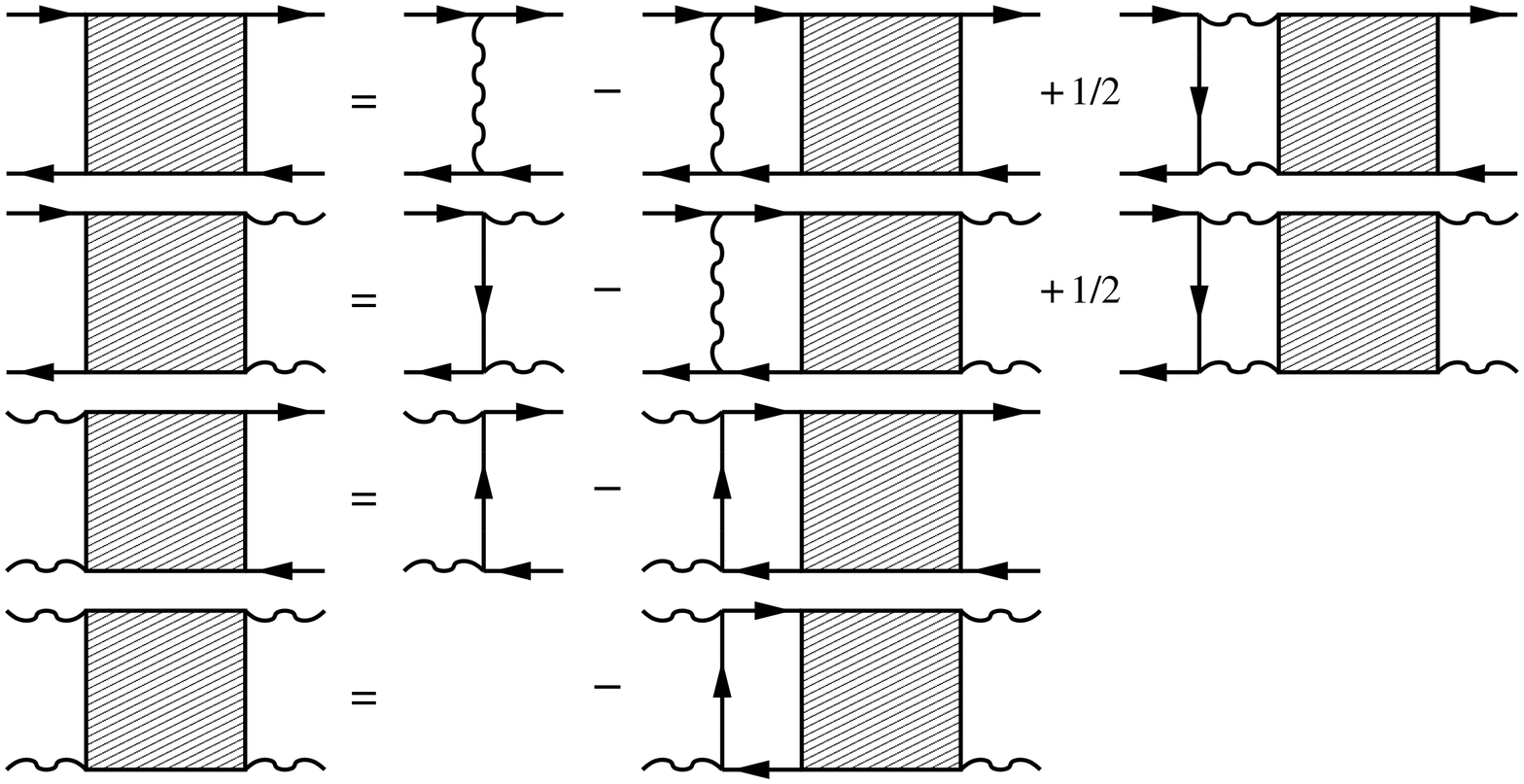,height=4.1cm}
\end{center}
 \vspace{-0.5cm}
 \caption{Integral equations for the 4-point functions in QED
at 2-loop order in the loop or at NLO in the~$1/N$~expansion. 
}
 \label{figintegralQED}
\end{figure}

\begin{figure}[b]
 \begin{center}
 \scalebox{0.6}[0.6]{\includegraphics{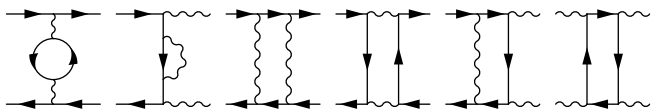}}
 \end{center}
 \vspace{-0.5cm}
 \caption{One-loop perturbative kernel in the 2-loop approximation in QED. 
 In the $1/N$ expansion, the first and the fourth diagrams dominate.}
 \label{figkernelQEDJose}
\end{figure}

In the loop expansion (see Fig.\ \ref{figgammaQED}), we restrict the
discussion to the 2-loop approximation. The corresponding integral
equations are shown in Fig.\ \ref{figintegralQED}.  In order to analyze
what processes are included in the weak-coupling limit, we follow the
strategy of the scalar theory and rewrite the integral equation such that
the external momenta are on-shell. The possible one-loop contributions to
the kernel are shown in Fig.\ \ref{figkernelQEDJose}. This kernel contains
the sum of the squares of all 2-to-2 processes, but not the interference
terms. Indeed, it is the 3-loop diagram that is responsible for
interference. Using the results
from the diagrammatic analysis \cite{ValleBasagoiti:2002ir}, we
find that, in order to determine the shear viscosity and the electrical
conductivity to leading-log order, only the first two rungs in Fig.\
\ref{figkernelQEDJose} need to be considered, with the momentum flowing
through the rung being soft and below the lightcone.  We conclude that the
integral equations in the 2-loop approximation sum the necessary
diagrams \cite{ValleBasagoiti:2002ir,Aarts:2002tn} to obtain the
shear viscosity and electrical conductivity in QED to leading-logarithmic
order.

Finally, we consider QED with $N$ identical fermions and a rescaled
coupling $e^2\to e^2/N$ in the large $N$ limit (see Fig.\
\ref{figgammaQED}). We consider only the NLO contribution.  In this
case the kernel is dominated by the first and the fourth diagrams in Fig.\
\ref{figkernelQEDJose}, the other contributions are suppressed by $1/N$.
The photon propagator still resums fermion bubbles and is in fact similar
to the auxiliary correlator in the $O(N)$ model.  Cutting the first and
fourth diagrams as in Fig.\ \ref{figkernelpert}, shows that they
correspond to Coulomb scattering in all three channels, which are indeed
\cite{Moore:2001fg} the dominant processes in a kinetic theory with many
fermions. We conclude that the 2PI-$1/N$ expansion in QED with $N$ 
fermions at NLO would yield the correct leading-order result for the shear 
viscosity and electrical conductivity.

In summary, we have shown how the calculation of transport coefficients is
organized in the framework of the 2PI effective action. For a variety of
models, we have discussed the first nontrivial truncations in a weak
coupling or $1/N$ expansion and found that these truncations yield ladder
integral equations with the particular kinematic configuration and rungs
appropriate to obtain transport coefficients at leading (logarithmic) order 
in the weak coupling or $1/N$ expansion. This formulation offers therefore a systematic
starting point for the derivation of these integral equations. Methods
for the actual solution of the integral equations (a topic not discussed
here) can be found in the literature.

In the wider context of nonequilibrium quantum field theory, our findings
provide theoretical support for the successful description of quantum
fields out of equilibrium using truncations of the 2PI effective action,
which are currently actively under investigation. We would like to emphasize 
that any scheme which aspires to properly describe the nonequilibrium evolution
and ensuing thermalization of a system has to render transport coefficients correctly.


\begin{acknowledgments}
{\em Acknowledgements.} Discussions with E.\ Braaten, U.\ Heinz, G.\ Moore, E.~Mottola and A.\ 
Schwenk are gratefully acknowledged.
 This work was supported by the DOE (Contract No.\ DE-FG02-01ER41190), the 
Basque Government and in part by the Spanish Science Ministry
(Grant FPA 2002-02037) and the University of the Basque Country (Grant
UPV00172.310-14497/2002). 
\end{acknowledgments}


\end{document}